          [2004/02/13 1.1 (PWD)]
\documentclass[a4paper,twocolumn]{esapub} % European paper

\usepackage{times}
\usepackage{natbib}
\usepackage{graphicx}

\title{The BATSE 9 year histories of the brightest AGN}
\author{A. B. Hill}
\author{E. J. Barlow}
\author{A. J. Bird}
\author{A. J. Dean}
\author{C. Ferguson}
\author{S. E. Shaw}
\author{M. J. Westmore}
\author{D. R. Willis}
\affil{School of Physics and Astronomy, University of Southampton, Hampshire, SO17 1BJ, United Kingdom}

\begin{document}


\maketitle

\begin{abstract}
The Burst and Transient Source Experiment (BATSE) aboard CGRO
monitored the whole sky through 8 NaI(Tl) crystal scintillators
sensitive in the 20 keV - 2 MeV energy band continuously from April
1991 until June 2000.  Results are presented on the long
term variabilty observed in the brightest Active Galactic Nuclei (AGN)
present in the BATSE data archive.  This was
achieved through the application of the Earth Occultation technique to
data flat-fielded using the BATSE Mass Model.  Removal of the temporal
background variations from the data should allow a more sensitive
extraction of source parameters.  Analysis of the
general trends of the 9-year light curves are also presented.
\end{abstract}

\section{Introduction}
The primary mission of the BATSE instrument was to detect and locate
$\gamma$-ray bursts.  While the 8 detectors had no intrinsic
positional sensitivity, burst positions were obtained by triangulation between
detectors.  It was realised, prior to launch, that persistent sources of $\gamma$-rays could also be observed
by using the Earth Occultation technique \citep{eot}.  This method
calculates the source flux by measuring the step in the count rate of
each BATSE detector as a source rises or sets below the Earth's limb.

Mass Modelling is a technique to simulate the background radiation
experienced by a space craft \citep{ssr}.  The BATSE Mass Model (BAMM)
is a GEANT3 Monte-Carlo simulation code developed at Southampton
\citep{alicante, bammpaper}.  BAMM simulates the expected count rate
from cosmic diffuse $\gamma$-rays, atmospheric albedo $\gamma$-rays
and cosmic rays thoughout the orbit.  The position and orientation of the
spacecraft relative to the Earth are used to filter these results and
generate the expected background.  BAMM has been used to
flat-field the entire 9 year BATSE data set.

It is thought that the origin of the exceptionally high luminosity
of Active Galaxies is the accretion of matter onto a supermassive
black hole located at the galaxy centre \citep{x-ray_vari}.
AGN typically have highly variable x-ray and $\gamma$-ray fluxes.  The
very short timescale variability indicates that this emission
originates from a small region very close to the central engine.
Hence the variability of AGN yields direct information about
the central supermassive black hole. 

\begin{figure*}[hbtp]
\centering
\includegraphics[width=0.78\linewidth]{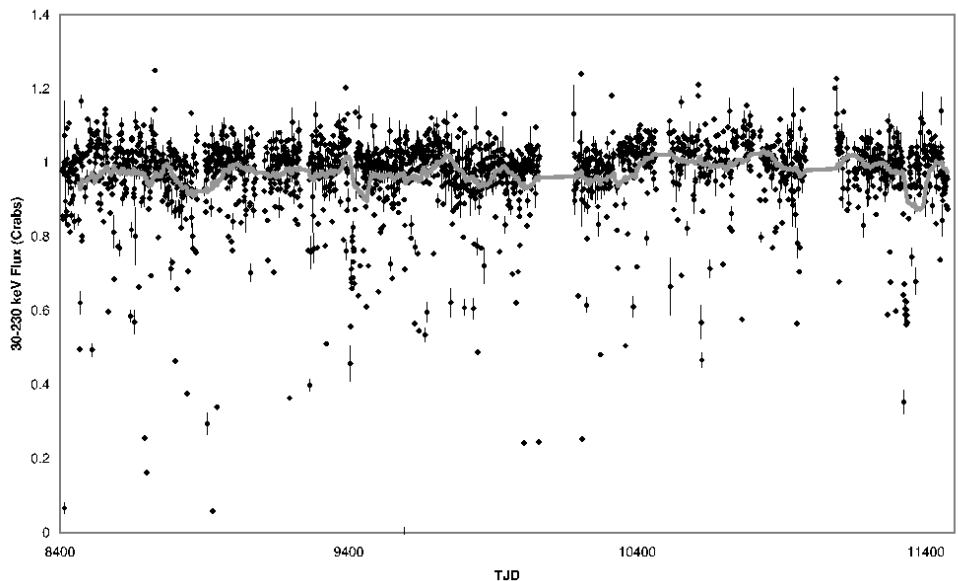}
\vspace{0.5cm}
\caption{The 9 year 30-230 keV light curve of the Crab Nebula as
measured on a daily basis.  The solid grey line is the 53 day (the
satellite precession period) moving average.\label{fig:crab}}
\end{figure*}

\section{Calibration using the Crab}
The Crab is one of the brightest $\gamma$-ray sources in the sky and
is well known for its lack of variability.  As such it is the ideal
`standard candle' with which to test the flat-fielded data set and
explore the capabilities of our methods.  Fig.~\ref{fig:crab} shows
the 9 year light curve of the Crab in the 30-230 keV energy
band as measured on a daily basis and with a 53 day moving average
fit; 53 days is the precession period of CGRO.

\begin{figure*}
\centering
\includegraphics[width=0.78\linewidth]{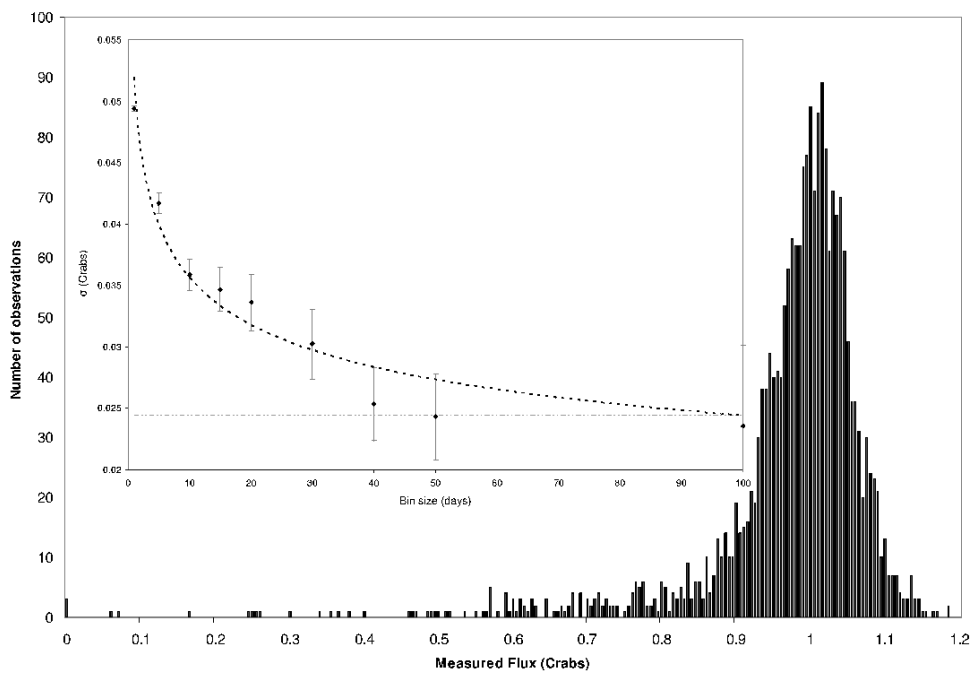}
\vspace{0.5cm}
\caption{The histogram of the Crab light curve seen in
Fig.~\ref{fig:crab}.  The inset graph indicates how the standard
deviation of the histogram decreases with increasingly large time
bins.  The broken black line indicates the power law decay.  The
broken grey line represents the asymptote which is indicative of the
systematic error.\label{fig:crabHist}}
\end{figure*}

A histogram of this light curve is seen in Fig.~\ref{fig:crabHist}.
The Crab appears relatively constant in the light curve, hence the
histogram should have
a Gaussian profile.  Fitting a Gaussian distribution to the histogram
yields a centroid of 1.005 $\pm$ 0.002 Crabs with a reduced $\chi$$^2$
of $\sim$3.  Examining the residuals
of this fit indicates a number of spurious points in the 0.6-0.9 Crab
region which may be the result of an unknown low level systematic,
removing these points renders a fit with a $\chi$$^2$ of 1.3.
Rebinning the data into longer timescale averages, before generating
the histogram, allows the estimation of the systematic component of
any errors.  As the time bins increase in size the standard deviation
of the data set would ideally asymptotically approach 0.  The inset
graph of Fig.~\ref{fig:crabHist} shows that the curve
approaches $\sim$2.5\%, indicating this level of systematic error.
This level is assumed in our studies of AGN light curves.

\begin{figure*}[htbp]
\centering
\includegraphics[totalheight=0.9\textheight, clip]{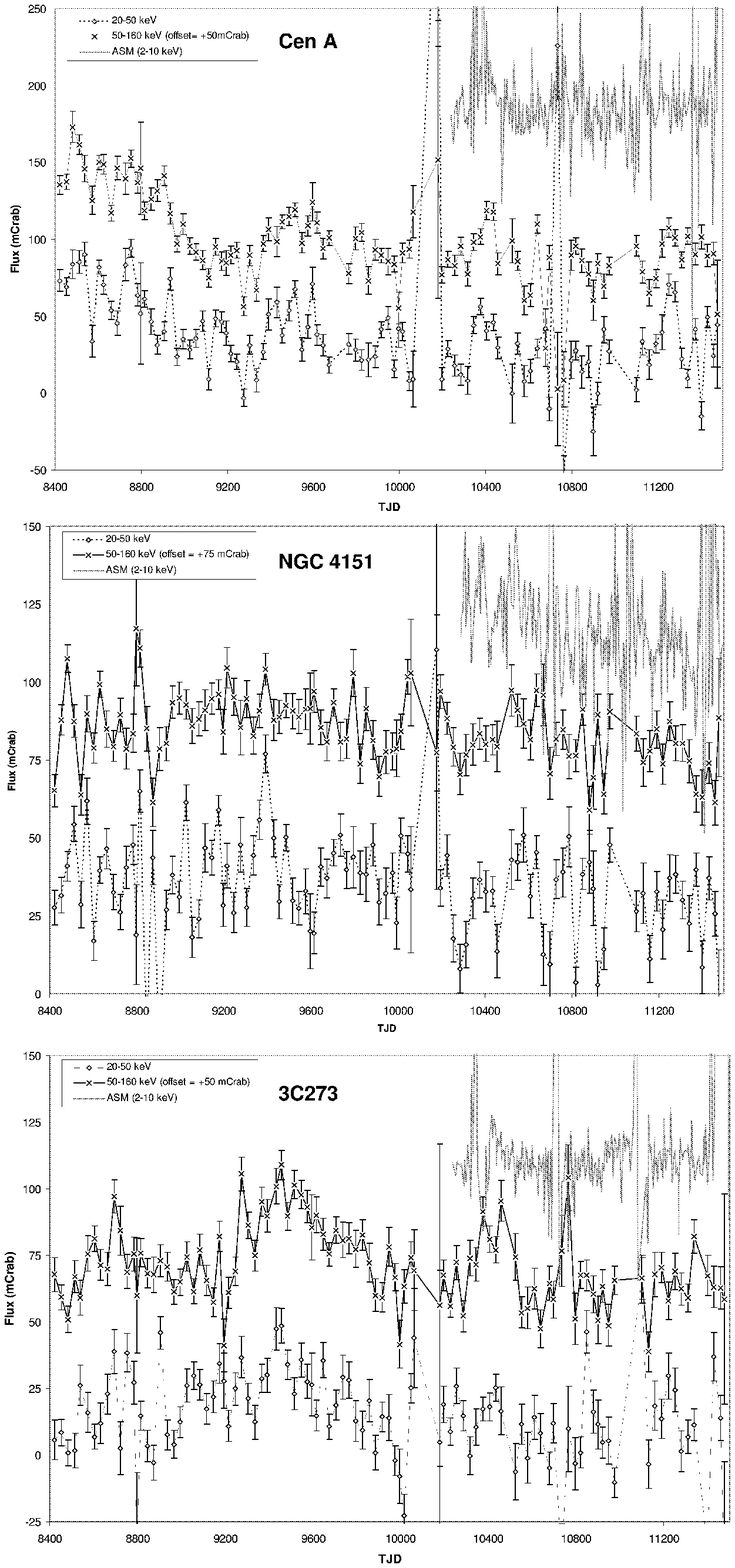}
\vspace{0.5cm}
\caption{The 9 year BATSE light curves of Cen A, NGC4151 and
3C273.  The broken black line with diamond points indicates the 20-50 keV flux; the solid black line with crosses indicates the offset 50-160 keV flux.  The solid dark grey
line is the offset 2-10 keV flux from the ASM on RXTE.\label{fig:lightcurves}}
\end{figure*}

\section{BATSE AGN}
\begin{figure}[htbp]
\centering
\includegraphics[width=0.95\linewidth,clip]{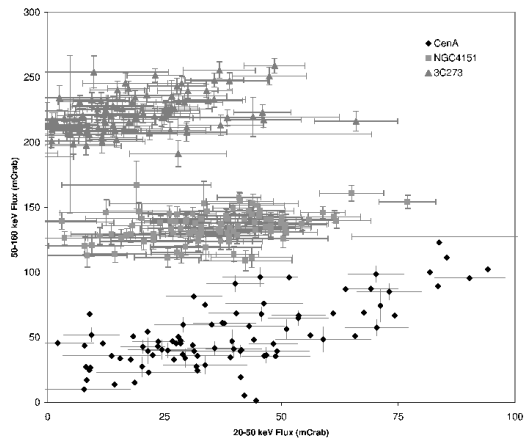}
\vspace{0.5cm}
\caption{Plotting the 50-160 keV (hard) flux against the 20-50 keV
(soft) flux of the AGN Cen A; NGC4151; 3C273.  NGC4151 and 3C273 hard
fluxes have been offset by +100 and +200 respectively for legibility.\label{fig:hard-soft}}
\end{figure}

AGN are some of the most powerful persistent
sources of $\gamma$-rays in the observable universe.  However the low
fluxes which arrive at Earth make them a very challenging subject to
observe in this waveband.  The three brightest AGN seen in BATSE are:
Centaurus A, a radio galaxy; NGC4151, a seyfert galaxy; 3C273, a
blazar.  Fig.~\ref{fig:lightcurves} shows the light curves in the
BATSE 20-50 keV and 50-160 keV band of these AGN.  The data have
been binned up to 30 day data points in order to minimise any errors.
Additionally the RXTE-ASM 2-10 keV light curves are presented, scaled
and offset from the BATSE curves to aid legibility.
Fig.~\ref{fig:hard-soft} is a hard flux - soft flux plot of the three
AGN showing the flux in the 30-50 keV band against that of the 50-160
keV band.

\subsection{Centaurus A}
The hard and soft BATSE light curves seen in the upper panel of
Fig.~\ref{fig:lightcurves} appear to be well correlated
apart from a short, $\sim$30 day outburst in the 20-50 keV band at TJD
$\sim$10800 which appears to coincide with a drop in the 50-160 keV flux.  However, the
20-50 keV flux continues to be correlated with the ASM light curve
during this period.  This appears to be corroborated by the hard - soft plot where the
hard and soft fluxes indicate a positive correlation.  A Spearman rank
correlation coefficient of 0.56 with a probability of chance occurance of p$<$0.0001 confirms this positive correlation.  The ASM light curve follows the general trends exhibited
by both of the BATSE light curves neglecting the outburst at TJD $\sim$10800.

\subsection{NGC 4151}
The hard and soft light curves shown in the middle panel of
Fig.~\ref{fig:lightcurves} appear to be well correlated although
there appears to be a potential time lag between them of $\sim$30
days.  Emission in the 50-160 keV band initially appears to precede that of the
20-50 keV band, however this is not obvious after TJD $\sim$8800.  As this lag is on the same time scale as the data
binning it is tenuous without a detailed statistical analysis.
The hard-soft plot looks flat as the hard flux appears independent of
the soft flux, however, a positive correlation is indicated by a Spearman rank
correlation coefficient of 0.31 with a probability of chance occurance of p$=$0.002.

\subsection{3C273}
Both light curves seen in the lower panel of
Fig.~\ref{fig:lightcurves} are synchronized and well correlated.  They
both show the same general trends.  The hard-soft plot indicates a
positive correlation with a Spearman rank
correlation coefficient of 0.45 with a probability of chance occurance of p$<$0.0001. 

\section{Discussion and Future Work}
The 9 year BATSE data set has been flat-fielded using BAMM to remove
temporal variations in the instrument background.
We have then applied the standard Earth Occultation Technique
developed at Marshall Space Flight Centre \citep{eot} on the
flat-fielded data set and made measurements to the Crab Nebula and 3
AGN: Cen A; NGC4151; 3C273.

The Crab light curve exhibits a constant flux of 1.005 $\pm$ 0.002 Crabs
confirming that the flat-fielding had no adverse effect on
the data.  Additionally, the systematic error of the methods
and analysis performed in the generation of the light curve is
estimated to be limited to $\sim$2.5\%.  The AGN light curves show
that BATSE is sensitive to the variations in their $\gamma$-ray fluxes.

Work is in progress to re-optimise the standard Earth Occultation
technique for use with the flat-fielded data with the intention of improving sensitivity and
precision of measurements.  We are simultaneously using the flat-fielded
data with new methods to generate all-sky images for
the whole 9 years \citep{basspaper, munich}.

\end{document}